\documentclass[11pt]{article}
\usepackage[a4paper,margin=1in]{geometry}
\usepackage[T1]{fontenc}
\usepackage[utf8]{inputenc}
\usepackage{lmodern}
\usepackage{microtype}
\usepackage{amsmath,amssymb,amsthm}
\usepackage{booktabs,longtable,array}
\usepackage[hyphens]{url}
\usepackage{xcolor}
\usepackage{hyperref}
\usepackage{setspace}
\hypersetup{
  colorlinks=true,
  linkcolor=black,
  citecolor=blue!50!black,
  urlcolor=blue!50!black,
  pdftitle={Preserving Decision Sovereignty in Military AI},
  pdfauthor={Peng WEI; Wesley Shu}
}
\title{Preserving Decision Sovereignty in Military AI:\\
A Trade-Secret-Safe Architectural Framework for Model Replaceability, Human Authority, and State Control}
\author{%
\begin{tabular}[t]{c@{\hspace{2.5cm}}c}
Peng WEI & Wesley Shu \\[0.3em]
College of Plant Protection & The Institute of Energetic Paradigm \\
Southwest University, China &  \\
\texttt{weipeng2019@swu.edu.cn} & \texttt{shuwesley@gmail.com}
\end{tabular}}
\date{March 2026}

\begin{document}
\maketitle

\begin{abstract}
Recent events surrounding the relationship between frontier AI suppliers and national-security customers have made a structural problem newly visible: once a privately governed model becomes embedded in military workflows, the supplier can influence not only technical performance but also the operational boundary conditions under which the system may be used. This paper argues that the central strategic issue is not merely access to capable models, but preservation of decision sovereignty: the state's ability to retain authority over decision policy, version control, fallback behavior, auditability, and final action approval even when analytical modules are sourced from commercial vendors. Using the public Anthropic--Pentagon dispute of 2026, the broader history of Project Maven, and recent U.S., NATO, U.K., and intelligence-community guidance as a motivating context, the paper develops a trade-secret-safe architectural formulation of the Energetic Paradigm as a layered, model-agnostic command-support design. In this formulation, supplier models remain replaceable analytical components, while routing, constraints, logging, escalation, and action authorization remain state-owned functions. The paper contributes three things: a definition of decision sovereignty for military AI; a threat model for supplier-induced boundary control; and a public architectural specification showing how model replaceability, human authority, and sovereign orchestration can reduce strategic dependency without requiring disclosure of proprietary implementation details. The argument is conceptual rather than experimental, but it yields concrete implications for procurement, governance, and alliance interoperability.
\end{abstract}

\section{Introduction}
Artificial intelligence is increasingly moving from back-office experimentation into operational, intelligence, and command-support environments. This shift is not occurring in a vacuum. In the United States, Project Maven made AI-enabled intelligence processing an institutional reality years before generative AI entered public consciousness \cite{mavenmemo,mssarmy}. By 2024 and 2025, frontier model vendors were openly partnering with defense-facing platforms and cloud providers to deliver large language models inside classified or highly controlled government environments \cite{palantiranthropic,awsreinvent,anthropicdod}. In NATO, allied acquisition of AI-enabled warfighting systems further demonstrated that AI is no longer merely a research topic; it is becoming part of military information and decision infrastructures \cite{nato_mss,nato_ai_2024}.

At the same time, the public dispute that emerged in March 2026 between Anthropic and the U.S. Department of Defense exposed a deeper structural tension. Public reporting indicates that the disagreement was not simply about model quality or price. It was about the extent to which a private supplier could preserve restrictions on military uses of its model, and the extent to which the government would tolerate those restrictions once the model had become operationally important \cite{reuters_lawsuit,reuters_background,wapo_lawsuit,ap_michael}. This dispute made a latent problem explicit: when military organizations depend on externally governed AI models, operational boundaries may no longer be set exclusively by the state.

This paper argues that the relevant strategic variable is therefore decision sovereignty. Military AI systems should be evaluated not only in terms of accuracy, latency, or scale, but also in terms of whether the sovereign user retains control over decision policy, versioning, routing, constraints, audit trails, and final action authorization. The paper develops a trade-secret-safe version of the Energetic Paradigm (EP) as a public architectural argument rather than a disclosure of proprietary mechanisms. The purpose is not to reveal implementation details. It is to make a narrow, defensible claim: military AI sovereignty is best preserved when analytical models are treated as replaceable modules inside a state-owned orchestration and assurance structure, rather than as the sole locus of system capability.

The contribution is conceptual rather than experimental. Specifically, the paper: (1) defines decision sovereignty for military AI; (2) identifies a threat model for supplier-induced boundary control; and (3) presents a model-agnostic, layered design in which final authority remains in sovereign structures rather than in vendor-governed models. This approach is consistent with responsible-AI guidance emphasizing accountability, traceability, governability, and human judgment in defense applications \cite{dod_ethics,dodrai,dodd300009,odni_ethics,ukmod_ai,nato_ai_2021,nato_ai_2024}. It is also responsive to wider concerns in the literature about trust calibration, human agency, accountability gaps, foundation-model dependence, and the political economy of AI supply chains \cite{mayer2023,bode2025,rajietal,krolletal,bommasani2021,brundage2018}.

\section{Motivating Context: From Vendor Access to Boundary Control}
The immediate context for this paper is the now-public conflict over military uses of frontier models. Anthropic's defense engagement was not peripheral. Public statements show that the company partnered with Palantir in late 2024 to bring Claude models to U.S. government intelligence and defense operations on AWS \cite{palantiranthropic}. AWS later highlighted the availability of Anthropic models in secure public-sector environments, including its Top Secret Region roadmap \cite{awsreinvent}. In July 2025, Anthropic announced a Department of Defense agreement with a ceiling of \$200 million to prototype frontier AI capabilities for national security \cite{anthropicdod}. Public system documentation and company materials further show that Anthropic had developed governance structures, usage policies, and system cards that explicitly shape how its models behave across sensitive domains \cite{anthropic_constitution,anthropic_syscard}.

The 2026 dispute revealed the political consequences of that governance architecture. According to Reuters, Washington Post, and Associated Press reporting, the Pentagon treated Anthropic as a supply-chain risk after disagreements over model guardrails, especially around autonomous weapons and domestic surveillance \cite{reuters_lawsuit,reuters_background,wapo_lawsuit,ap_michael}. Whether one agrees with Anthropic's position or the government's response is not the main point here. The more important point is structural: the dispute showed that a privately governed model can become strategically consequential enough that its safety boundaries are interpreted as constraints on state action.

This episode sits within a longer trajectory. Project Maven was created to convert enormous volumes of sensor data into actionable intelligence at speed \cite{mavenmemo}. Over time, military AI systems evolved from narrowly scoped computer vision tools into broader command-support ecosystems, including data integration, targeting support, and planning assistance \cite{mssarmy,probasco2025,nurkin2023}. NATO's acquisition of the Maven Smart System and broader alliance work on AI governance indicate that these capabilities are diffusing institutionally, not merely experimentally \cite{nato_mss,nato_ai_2024,lucarelli2021}. The result is a new class of dependency: not just reliance on outside hardware or code, but reliance on externally governed inference services whose policies can shape action space.

The literature on military AI has described many adjacent concerns: acceleration of tempo, opacity, bias, trust calibration, accountability, and the risk that decision support may drift into decision delegation \cite{nadibaidze2024,trabucco2025,mayer2023,bode2025,mcfarland2022}. What has received less explicit treatment is supplier boundary control: the risk that a commercial vendor's policies, update decisions, or service withdrawal can alter the practical decision space of a sovereign military organization. Public-sector AI acquisition guidance increasingly recognizes related problems by stressing version control, rollback rights, vendor documentation, and evaluation before deployment of new model versions \cite{omb_m2418,omb_m2522}. However, these procurement remedies do not by themselves solve the architectural issue. If the system's core logic is concentrated in an external supplier model, then the sovereign actor may still remain dependent on the supplier's continuing policy acceptance.

\section{Problem Definition: Decision Sovereignty}
For the purposes of this paper, decision sovereignty is defined as the capacity of a state or authorized military institution to retain authoritative control over the following six elements of an AI-supported decision process:

\begin{enumerate}
\item \textbf{Policy sovereignty}: control over what kinds of outputs may inform which operational actions.
\item \textbf{Routing sovereignty}: control over how tasks are allocated across analytical modules, fallbacks, and review channels.
\item \textbf{Version sovereignty}: control over model substitution, rollback, pinning, and upgrade timing.
\item \textbf{Constraint sovereignty}: control over refusal logic, escalation thresholds, and use-case boundaries.
\item \textbf{Audit sovereignty}: control over logging, provenance, explainability records, and reviewability.
\item \textbf{Action sovereignty}: control over final approval for recommendation adoption, especially where lethal or coercive effects are implicated.
\end{enumerate}

This definition builds on but is distinct from common discussions of human control. A system can keep a human nominally in the loop while still lacking decision sovereignty if the available options, model behavior, or operational boundaries are effectively determined by an external supplier. Conversely, a system may preserve decision sovereignty when advanced external models are used extensively, provided they remain subordinate to a sovereign orchestration layer that governs routing, constraints, logging, and final action approval.

This distinction matters because military AI is not merely a technical tool. It is part of a chain that connects perception, interpretation, prioritization, recommendation, and action. If the system's most consequential boundary conditions are governed externally, then the state's command authority is diluted even if operators remain formally responsible.

\section{Threat Model: Supplier-Induced Boundary Control}
The paper identifies five pathways by which commercial AI suppliers can shape sovereign military decision space without possessing formal command authority.

\subsection{Policy Injection}
Suppliers can encode use restrictions, refusals, and content boundaries into model behavior or service terms. In commercial settings this may be normal governance. In military settings it can amount to practical boundary control if the model is deeply integrated into operational analysis.

\subsection{Version Drift}
A provider can update the model, safety stack, moderation logic, or access conditions. Even beneficial updates may alter the recommendation surface, error profile, or willingness to process certain requests. If the sovereign user lacks strong pinning and rollback rights, this becomes a form of dependency.

\subsection{Withdrawal Leverage}
Access can be suspended, narrowed, repriced, or delayed. The possibility of withdrawal itself changes bargaining power, particularly when an institution has built workflows around one model family.

\subsection{Audit Asymmetry}
If key system behavior is visible only through provider documentation, black-box APIs, or system cards, the operator may be unable to reconstruct why certain outputs were suppressed, transformed, or refused in a critical context.

\subsection{Normative Drift Through Convenience}
Over time, operators and procurement teams may adapt their own expectations to the affordances and limits of a dominant supplier model. This produces a softer but still consequential dependency: institutional practice begins to mirror vendor governance.

These pathways do not require malign intent. They arise from ordinary commercial governance interacting with extraordinary military responsibility. The issue is structural, not psychological.

\section{Trade-Secret-Safe EP Architecture}
The original Energetic Paradigm contains internal concepts and methods that are not appropriate for public disclosure. This paper therefore presents only a trade-secret-safe architectural specification. The purpose is to describe what a sovereignty-preserving design must do, not how proprietary internal mechanisms are implemented.

The public EP architecture can be expressed as a layered command-support framework:

\begin{enumerate}
\item \textbf{Input and provenance layer}: ingests data, classifies source reliability, attaches provenance, and records uncertainty.
\item \textbf{Analytical model layer}: one or more replaceable external or internal AI models generate summaries, anomaly flags, option sets, or structured analytical outputs.
\item \textbf{Sovereign orchestration layer}: a state-controlled logic layer handles routing, fallback selection, confidence gates, escalation rules, and interface normalization across models.
\item \textbf{Constraint and review layer}: applies policy-specific checks, human review requirements, red-team triggers, and domain restrictions that are owned by the sovereign operator rather than by the vendor.
\item \textbf{Authorization layer}: ensures that operationally consequential actions cannot proceed on model output alone and require designated human or institutional approval.
\item \textbf{Audit and rollback layer}: logs model choice, version, prompts, context boundaries, rule triggers, human interventions, and action outcomes for ex post review and rapid rollback.
\end{enumerate}

The key principle is modular asymmetry. Models may be commercially supplied and may even be best-in-class. But they are not allowed to become the system's final arbiter. The intelligence or recommendation value of a model is separated from the sovereign control function that determines whether, where, and how that value enters action processes.

\section{Architectural Principle: Replaceability Over Dependence}
A sovereignty-preserving system should be designed around replaceability rather than model loyalty. In this context, replaceability means that the system can substitute one analytical model for another, or operate in a degraded fallback mode, without losing policy continuity, audit continuity, or human authority.

This principle has several consequences.

First, interfaces should normalize outputs into a common schema before they enter the sovereign orchestration layer. That prevents downstream command logic from becoming entangled with the idiosyncrasies of one provider.

Second, critical policy functions such as escalation triggers, red lines, and authorization thresholds should not be embedded solely in vendor prompts, hidden safety layers, or provider-managed moderation systems. Those functions belong in the sovereign layer.

Third, operational continuity must include degraded modes. A sovereignty-preserving architecture must specify what happens when a model is unavailable, restricted, or uncertified for a given context. A system that fails closed in all such cases may be safer than one that fails open, but it is still strategically brittle if there is no viable replacement path.

Fourth, replaceability changes the procurement logic. The goal is not to eliminate commercial dependence altogether; that may be unrealistic. The goal is to prevent dependence from maturing into decision capture.

\section{Human Authority as a System Property}
Many discussions of military AI invoke ``human in the loop'' as a sufficient safeguard. That phrase is too thin. Human authority is not preserved simply because a person is present somewhere in the chain. It is preserved when the architecture ensures that model outputs remain advisory, contestable, attributable, and interruptible.

A human authority-preserving system should therefore have at least four properties:

\begin{enumerate}
\item \textbf{Contestability}: authorized personnel can reject or override model outputs without breaking workflow integrity.
\item \textbf{Traceability}: personnel can see what model, version, data provenance, and rules shaped the output.
\item \textbf{Escalation structure}: ambiguous or high-consequence cases automatically route to higher levels of review.
\item \textbf{Non-self-execution}: model output cannot directly trigger kinetic, coercive, or strategically consequential action.
\end{enumerate}

This goes beyond ethics slogans. It makes human authority a property of system design. In other words, authority is preserved not by trusting the operator to behave cautiously under pressure, but by embedding review, contestation, and interruption into the architecture itself.

\section{Public-Safe Operational Logic}
Because this paper is trade-secret safe, it does not expose internal algorithms or proprietary mechanisms. But the public-safe operational logic can still be expressed in high-level form.

A sovereignty-preserving military AI workflow proceeds as follows:

\begin{enumerate}
\item Data and task requests enter through controlled interfaces with provenance tagging.
\item One or more analytical models produce structured outputs, not executable commands.
\item Sovereign orchestration evaluates confidence, context, domain restrictions, and fallback conditions.
\item Constraint checks determine whether the output is admissible for further consideration.
\item Human reviewers inspect recommendations at policy-defined checkpoints.
\item Only authorized institutional actors may convert reviewed outputs into operational decisions.
\item Full logs support audit, investigation, rollback, and post-action learning.
\end{enumerate}

The importance of this sequence is not procedural formality. It is placement of the decision boundary. The decisive transition from analysis to action occurs after sovereign review, not inside the supplier-governed model.

\section{Comparison with Model-Centric Integration}
To clarify the claim, it helps to contrast two stylized architectures.

In a \emph{model-centric integration}, a military organization connects its workflow directly to a frontier model or to a provider-controlled stack. The model performs analysis, mediation, and boundary enforcement. Logging and operational interpretation may be partly available, but the provider's update cadence, refusal logic, and acceptable-use interpretation remain structurally central.

In a \emph{sovereignty-centric integration}, frontier models still provide advanced analysis, but they do so behind a state-owned orchestration, assurance, and authorization structure. The model is important, but not decisive in the constitutional sense. The decisive element is the architecture that governs model use.

Table~\ref{tab:comparison} summarizes the distinction.

\begin{table}[h]
\centering
\caption{Model-centric vs. sovereignty-centric military AI integration}
\label{tab:comparison}
\begin{tabular}{p{0.28\linewidth}p{0.30\linewidth}p{0.30\linewidth}}
\toprule
Dimension & Model-centric integration & Sovereignty-centric integration \\
\midrule
Primary source of operational boundary & Supplier model policies and service conditions influence behavior & State-owned orchestration and policy layers govern admissibility and action \\
Model substitution & Difficult if workflow is deeply coupled to one provider & Designed for replaceability across providers or internal models \\
Failure mode under vendor withdrawal & Operational degradation may be severe & Structured substitution and degraded-mode operation possible \\
Version governance & Provider cadence dominates unless separately negotiated & Version pinning, rollback, and replacement embedded in architecture \\
Auditability & Partially dependent on vendor disclosure & Logging and escalation governed by operator \\
Human authority & Vulnerable to automation drift under speed pressure & Explicit checkpoints preserved at policy-defined stages \\
Alliance interoperability & Can depend on common vendor acceptance & Can use different vendor modules behind shared policy interface \\
Strategic sovereignty & Reduced when model access defines action space & Preserved when models remain analytical components \\
\bottomrule
\end{tabular}
\end{table}

This distinction does not imply that sovereign architectures reject frontier models. On the contrary, the point is to use them more safely and more strategically. A capable external model can be extremely valuable for summarization, anomaly detection, simulation, or option generation. What should be avoided is concentration of command relevance in a component whose policies and lifecycle remain under outside control.

\section{Implications for Procurement, Governance, and Alliance Use}
The framework has several practical implications.

First, procurement should evaluate \emph{replaceability} as a core performance criterion. Current public-sector acquisition guidance already moves in this direction by emphasizing documentation, evaluation, rollback provisions, and treatment of models as versioned subcomponents \cite{omb_m2418,omb_m2522}. For military AI, this should be extended into a sovereignty test: can the mission continue if one supplier changes policy, withdraws service, or fails certification?

Second, responsible AI in defense should be interpreted architecturally, not merely normatively. High-level principles such as responsibility, traceability, and governability matter only insofar as they are instantiated in system design \cite{dod_ethics,dodrai,nato_ai_2024,ukmod_ai,papagiannidis2025}. A military organization that depends wholly on one external model may satisfy an ethics checklist while still lacking sovereign control.

Third, alliance interoperability may benefit from policy-layer standardization rather than model-layer standardization. NATO's revised AI strategy stresses TEVV, responsible use, and interoperable adoption \cite{nato_ai_2024}. A sovereignty-preserving architecture suggests that allies do not need identical models. They need compatible interfaces for provenance, constraints, audit logging, and human authorization.

Fourth, the framework narrows the trust problem. Literature on human-autonomy teaming shows that the key issue is not whether operators should ``trust AI'' in the abstract, but how trust should be calibrated to role, context, and uncertainty \cite{mayer2023,mcfarland2022,bode2025}. If analytical modules are replaceable and non-authoritative, trust can be distributed and bounded, rather than concentrated in one opaque provider-controlled engine.

Fifth, this architecture is compatible with zero-trust and high-assurance security approaches to large language model deployment. Recent guidance on LLM-based systems emphasizes traceability, compartmentalization, and continuous supervision \cite{bsi_anassi}. Those ideas map naturally onto a military decision-sovereignty framework in which orchestration and assurance remain sovereign assets.

\section{Relation to Existing AI Governance Debates}
The argument developed here is narrower than many broad debates about AI ethics, but it intersects with them. Work on robust and beneficial AI, AI safety, and malicious use has long warned that powerful AI systems create novel governance problems when incentives, control, and accountability are poorly aligned \cite{russell2015,amodei2016,brundage2018}. Research on foundation models further stresses that highly general systems create downstream dependency, opacity, and concentration risks that extend beyond any single application \cite{bommasani2021,weidinger2022}. Studies of accountability gaps and algorithmic due process show that delegating consequential functions to opaque systems can undermine legal and institutional responsibility \cite{krolletal,rajietal}.

Military contexts intensify these concerns because the costs of failure are unusually high and because the public-private boundary is structurally central. The state remains legally accountable for military action, but increasingly relies on commercial actors for compute, models, data infrastructure, and interfaces. That is why decision sovereignty deserves explicit treatment as a distinct design principle. It translates general governance concerns into a specific institutional question: where, concretely, does the decision boundary live?

In this sense, the proposed framework is also compatible with the emerging emphasis on meaningful human control and human responsibility in military AI debates \cite{hrw300009,taddeo2021,mcfarland2022}. The key contribution here is to show that meaningful human control is not only a matter of keeping a person ``in the loop.'' It also depends on preventing supplier-controlled systems from becoming the de facto source of operational boundary-setting. Human oversight becomes fragile when the menu of machine-generated options, refusals, and defaults is itself externally governed.

\section{Limitations and Future Research}
This paper is intentionally limited. It is a conceptual and architectural argument, not an empirical validation study. It does not publish internal EP methods, benchmark data, targeting logic, or deployment playbooks. It does not argue for autonomous lethal decision-making, nor does it claim that architectural redesign eliminates all risks associated with military AI.

Several research tasks remain open. One is the development of empirical tests for decision sovereignty, including measures of substitution resilience, policy-layer stability, and audit completeness across vendor changes. A second is legal analysis of how sovereign orchestration interacts with alliance rules, procurement law, and accountability in multinational operations. A third is organizational: military institutions will need training, doctrine, and staffing models capable of treating orchestration, assurance, and audit as command functions rather than as back-end IT issues \cite{lucarelli2021,nurkin2023,pfaff2023}. A fourth concerns evaluation under contested conditions, including degraded communications, adversarial deception, and cloud disruption.

Future public work could also compare architecture choices across allied systems and assess whether procurement standards should explicitly require model replaceability and sovereign policy ownership. Such work would help move the discussion from abstract ethics to verifiable system properties.

\section{Conclusion}
The strategic question raised by frontier AI in military settings is not only who has the best model. It is who retains control when models become part of the decision infrastructure. The public Anthropic--Pentagon dispute of 2026 made visible a structural reality that had been building for years: as commercial models enter military workflows, supplier governance can begin to shape sovereign action space \cite{reuters_lawsuit,reuters_background,wapo_lawsuit}. That is a strategic design problem, not merely a contractual inconvenience.

This paper has argued that decision sovereignty provides the right lens for analyzing the problem. A military AI system is more sovereign when state-owned policy, routing, logging, and authorization layers remain in command, while supplier models remain replaceable analytical modules. The trade-secret-safe public formulation of the Energetic Paradigm presented here does not disclose proprietary internals. It makes a narrower and more defensible claim: military AI sovereignty depends less on ownership of every model than on ownership of the architecture that determines how models are used, constrained, substituted, and overridden.

If that claim is correct, then future military advantage will depend not only on access to frontier AI capability, but on the institutional capacity to embed such capability without surrendering the decision boundary itself.

\end{document}